\documentclass{emulateapj}
\usepackage{apjfonts}
\usepackage{graphicx}
\usepackage{xspace}

\begin{document}

\lefthead{The Fast X-ray Transient NuSTAR J163433--4738.7}
\righthead{Tomsick et al.}

\submitted{Accepted by ApJ}

\def\lsim{\mathrel{\lower .85ex\hbox{\rlap{$\sim$}\raise
.95ex\hbox{$<$} }}}
\def\gsim{\mathrel{\lower .80ex\hbox{\rlap{$\sim$}\raise
.90ex\hbox{$>$} }}}

\title{NuSTAR J163433--4738.7: A Fast X-ray Transient in the Galactic Plane}

\author{John A. Tomsick\altaffilmark{1},
Eric V. Gotthelf\altaffilmark{2},
Farid Rahoui\altaffilmark{3,4},
Roberto J. Assef\altaffilmark{5},
Franz E. Bauer\altaffilmark{6,7},
Arash Bodaghee\altaffilmark{1},
Steven E. Boggs\altaffilmark{1},
Finn E. Christensen\altaffilmark{8},
William W. Craig\altaffilmark{1,9},
Francesca M. Fornasini\altaffilmark{1,10},
Jonathan Grindlay\altaffilmark{11},
Charles J. Hailey\altaffilmark{2},
Fiona A. Harrison\altaffilmark{12},
Roman Krivonos\altaffilmark{1},
Lorenzo Natalucci\altaffilmark{13},
Daniel Stern\altaffilmark{14},
William W. Zhang\altaffilmark{15}}

\altaffiltext{1}{Space Sciences Laboratory, 7 Gauss Way, University of California, Berkeley, 
CA 94720-7450, USA; jtomsick@ssl.berkeley.edu}

\altaffiltext{2}{Columbia Astrophysics Laboratory, Columbia University, New York, NY 10027, USA}

\altaffiltext{3}{European Southern Observatory, Karl Schwarzschild-Strasse 2, 85748 Garching bei 
M{\"u}nchen, Germany}

\altaffiltext{4}{Department of Astronomy, Harvard University, 60 Garden Street,
Cambridge, MA 02138, USA}

\altaffiltext{5}{N{\'u}cleo de Astronom{\'i}a de la Facultad de Ingenier{\'i}a, Universidad Diego
Portales, Av. Ej{\'e}rcito 441, Santiago, Chile}

\altaffiltext{6}{Instituto de Astrof{\'i}sica, Facultad de F{\'i}sica, Pontifica Universidad
Cat{\'o}lica de Chile, 306, Santiago 22, Chile}

\altaffiltext{7}{Space Science Institute, 4750 Walnut Street, Suite 205, Boulder, CO 80301, USA}

\altaffiltext{8}{DTU Space, National Space Institute, Technical University of Denmark, Elektrovej 
327, DK-2800 Lyngby, Denmark}

\altaffiltext{9}{Lawrence Livermore National Laboratory, Livermore, CA 94550, USA}

\altaffiltext{10}{Astronomy Department, University of California, 601 Campbell Hall, Berkeley, 
CA 94720, USA}

\altaffiltext{11}{Harvard-Smithsonian Center for Astrophysics, Cambridge, MA 02138, USA}

\altaffiltext{12}{Cahill Center for Astronomy and Astrophysics, California Institute of 
Technology, Pasadena, CA 91125, USA}

\altaffiltext{13}{Istituto Nazionale di Astrofisica, INAF-IAPS, via del Fosso del Cavaliere, 00133 Roma, Italy}

\altaffiltext{14}{Jet Propulsion Laboratory, California Institute of Technology, Pasadena, CA 91109, USA}

\altaffiltext{15}{NASA Goddard Space Flight Center, Greenbelt, MD 20771}

\begin{abstract}

During hard X-ray observations of the Norma spiral arm region by the {\em Nuclear 
Spectroscopic Telescope Array (NuSTAR)} in 2013 February, a new transient source, 
NuSTAR~J163433--4738.7, was detected at a significance level of 8-$\sigma$ in 
the 3--10\,keV bandpass.  The source is consistent with having a constant {\em NuSTAR} 
count rate over a period of 40\,ks and is also detected simultaneously by {\em Swift} 
at lower significance.  The source is not significantly detected by {\em NuSTAR}, 
{\em Swift}, or {\em Chandra} in the days before or weeks after the discovery of the 
transient, indicating that the strong X-ray activity lasted for between $\sim$0.5 and 
1.5 days.  Near-IR imaging observations were carried out before and after the X-ray 
activity, but we are not able to identify the counterpart.  The combined {\em NuSTAR} 
and {\em Swift} energy spectrum is consistent with a power-law with a photon index of 
$\Gamma = 4.1^{+1.5}_{-1.0}$ (90\% confidence errors), a blackbody with 
$kT = 1.2\pm 0.3$\,keV, or a bremsstrahlung model with $kT = 3.0^{+2.1}_{-1.2}$\,keV.  
The reduced-$\chi^{2}$ values for the three models are not significantly different, 
ranging from 1.23 to 1.44 for 8 degrees of freedom.  The spectrum is strongly
absorbed with $N_{\rm H} = (2.8^{+2.3}_{-1.4})\times 10^{23}$\,cm$^{-2}$, 
$(9^{+15}_{-7})\times 10^{22}$\,cm$^{-2}$, and $(1.7^{+1.7}_{-0.9})\times 10^{23}$\,cm$^{-2}$,
for the power-law, blackbody, and bremsstrahlung models, respectively.  Although the 
high column density could be due to material local to the source, it is consistent with 
absorption from interstellar material along the line of sight at a distance of 11\,kpc, 
which would indicate an X-ray luminosity $>$$10^{34}$\,erg~s$^{-1}$.  Although we do not 
reach a definitive determination of the nature of NuSTAR~J163433--4738.7, we suggest 
that it may be an unusually bright active binary or a magnetar.

\end{abstract}


\section{Introduction}

Hard X-ray surveys of the Galaxy provide an opportunity to discover populations 
of extreme sources.  The promise of such surveys has been partially realized with the 
{\em International Gamma-Ray Astrophysics Laboratory} \citep[{\em INTEGRAL},][]{winkler03},
which carried out a 20--100\,keV survey of the entire Galactic Plane and has discovered 
hundreds of new sources \citep{bird10,krivonos12}, including new types of High Mass
X-ray Binaries (HMXBs), pulsar wind nebulae (PWNe), and magnetic Cataclysmic Variables 
(CVs).  The {\em Nuclear Spectroscopic Telescope Array} \citep[{\em NuSTAR},][]{harrison13}, 
which launched in 2012 June and covers the 3--79\,keV bandpass, is the first focusing hard 
X-ray telescope in orbit.  While it has a much smaller field of view than {\em INTEGRAL}, 
it has much lower background and greatly improved sensitivity.  Thus, one of {\em NuSTAR}'s
science goals is to extend our view deeper into the Galactic plane to look for hidden 
hard X-ray populations.  

During its first year of operation, {\em NuSTAR} has initiated surveys of $\sim$1 deg$^{2}$
areas both in the Galactic Center \citep[e.g.,][]{mori13,nynka13} and in a region which 
samples the spiral arm population in order to probe potentially 
different environments where X-ray binaries are found.  A spiral arm region centered on
Galactic coordinates of $l = 337.5^{\circ}$ and $b = 0^{\circ}$ was chosen for having the highest 
known density of OB star associations \citep{russeil03} and HMXBs \citep{bodaghee07,bodaghee12}.  
This is part of the ``Norma'' spiral arm region, which was identified early in the 
{\em INTEGRAL} mission as having an unusually high density of hard X-ray sources 
\citep{tomsick04_munich,dean05,lutovinov05a}.  The combination of active star formation 
and evidence that compact objects have already formed suggests that a survey by {\em NuSTAR} 
may uncover compact objects associated with populations of massive stars such as magnetars 
or faint HMXBs that are early in their evolutionary process and may have neutron star or 
black hole accretors.

The full surveys of the Galactic Center and the Norma region will be carried out 
over a period of $\sim$2 years; here we report on the discovery of a transient source 
made during the first part of the survey.  In the following, \S\,2 describes observations 
made with {\em NuSTAR}, {\em Swift}, and {\em Chandra}, as well as 
the procedures we used to reduce the data.  The results are presented in \S\,3, and we
discuss possibilities for the nature of the transient in \S\,4.

\section{Observations and Data Reduction}

As the first part of the {\em NuSTAR} survey of the Norma region, nine $\sim$20\,ks 
{\em NuSTAR} observations were performed between UT 2013 February 20 and February 24.  
Each $13^{\prime}$-by-$13^{\prime}$ field of view was partially overlapping with adjacent 
pointings, and the entire region covered was $\sim$0.2 deg$^{2}$.  The results from all 
nine pointings will be reported in Bodaghee et al. (submitted).  Here we focus on 
the observations that covered a new transient, NuSTAR~J163433--4738.7.  These 
observations are listed in Table~\ref{tab:obs}, including two that were obtained 
during the survey and a follow-up {\em NuSTAR} observation on 2013 March 23 that was 
coordinated with {\em Chandra}.

In addition, $\sim$2\,ks {\em Swift} X-ray Telescope (XRT) observations of the region 
were carried out during 2013 February 21-24, and four of these observations covered 
NuSTAR~J163433--4738.7.  Table~\ref{tab:obs} lists these along with three other 
{\em Swift} observations that were acquired after our survey as part of another 
observing program.  We also analyzed archival data covering the source, including 
two {\em Chandra} observations (ObsIDs 12529 and 12532 with exposure times of 
19.0\,ks and 19.5\,ks, respectively) that were acquired in 2011 as part of a survey 
of the same region being covered by {\em NuSTAR} (Fornasini et al., submitted).  

We reduced the {\em NuSTAR} and {\em Swift} data using HEASOFT v6.14 and the 
latest version of the Calibration Database (CALDB) files as of 2013 August 30.
We produced cleaned event lists for the {\em NuSTAR} Focal Plane Modules 
(FPMA and FPMB) using {\ttfamily nupipeline} and for the {\em Swift}/XRT using 
{\ttfamily xrtpipeline}, and further analysis of the event lists is described below.  
For {\em Chandra}, we processed the Advanced CCD Imaging Spectrometer 
\citep[ACIS,][]{garmire03} data with the {\em Chandra} Interactive Analysis of 
Observations (CIAO) software, using {\ttfamily chandra\_repro} to make event lists.

We obtained near-IR observations covering the NuSTAR~J163433--4738.7 error region.  
This includes $J$, $H$ and $K_{s}$ observations performed on 2011 July 19 with 
CTIO/NEWFIRM in the framework of near-IR mapping of the {\em Chandra} survey field. 
A detailed description of the data and their reduction can be found in Rahoui et al. 
(in prep).  They were reduced with the dedicated IRAF package \textsc{nfextern} 
following the standard procedure -- tailored for wide-field mosaics -- which consists 
of bad pixel removal, dark subtraction, linearity correction, flatfielding and median 
sky subtraction. The resulting images were then flux-calibrated through relative 
photometry with the 2MASS catalogue.  We also obtained $K_{s}$-band imaging with the 
Ohio-State Infra-Red Imager/Spectrometer (OSIRIS) at the Southern Astrophysical Research 
(SOAR) 4.1m Telescope.  We used the f/7 camera, providing a $80^{\prime\prime}$ field of 
view, centered at the nominal coordinates of NuSTAR~J163433-4738.7. We obtained 36$\times$60\,s 
dithered exposures of the field on 2013 April 3 under good conditions with seeing of 
$0.\!^{\prime\prime}7$, and observed it for 9$\times$60\,s again on 2013 April 5, also 
under good conditions but with somewhat worse seeing of $1.\!^{\prime\prime}0$.  Reductions 
were done using the {\ttfamily XDIMSUM} IRAF package and photometric calibration was 
obtained by comparing to 2MASS All-Sky Point Source Catalog sources in the field.

\section{Results}

\begin{figure}
\includegraphics[scale=0.47]{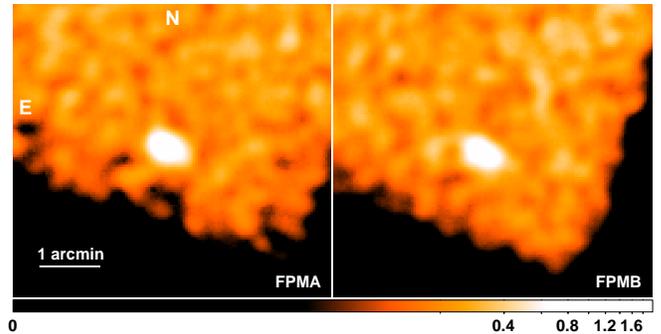}
\caption{\small {\em NuSTAR} discovery images in the 3--20\,keV energy band for 
NuSTAR~J163433--4738.7.  The source was detected in both of the {\em NuSTAR} 
Focal Plane Modules (FPMA and FPMB) in a 22.6\,ks exposure taken on 2013 Feb. 23-24.  
The images have been rebinned so that the pixel size is $2.\!^{\prime\prime}5$
and smoothed with a 6-pixel Gaussian.  The scale on the bottom of the figure is
in counts per pixel, and a logarithmic scaling is used.  The apparent elongation 
of the source is due to the distorted PSF shape at large off-axis angles.
\label{fig:nuimages}}
\end{figure}

The new transient was discovered from an inspection of the image from 
the 22.6\,ks {\em NuSTAR} observation that took place starting on 
2013 February 23, 14.52 h.  As shown in Figure~\ref{fig:nuimages}, the
source was detected in FPMA and FPMB.  To determine the significance of 
the detection, we extracted 3--10\,keV counts from a $30^{\prime\prime}$-radius 
circle centered on the approximate position of the source.  We determined 
the background level using a nearby source-free circular region with a 
radius of $90^{\prime\prime}$.  After background subtraction, we obtained 
$97\pm 14$ and $99\pm 20$ counts in FPMA and FPMB, respectively.  The 
combined significance is 8.0-$\sigma$, confirming the detection.  

To determine the source position, we extracted all the events from a 
$60^{\prime\prime}$-by-$60^{\prime\prime}$ square region and made histograms
of the counts, binning in the R.A. and Decl. directions.  We performed 
$\chi^{2}$ fitting of the histograms with a model consisting of a constant 
(accounting for the flat background) and a Gaussian for the source.  It 
should be noted that this is an approximation since the {\em NuSTAR} point
spread function (PSF) is non-Gaussian.  We determined the centroids separately 
for FPMA and FPMB, and they are consistent with each other.  The weighted average 
of the two centroids is R.A.=$16^{\rm h}34^{\rm m}33.\!^{\rm s}42$,
Decl.=--$47^{\circ}38^{\prime}41.\!^{\prime\prime}9$ (J2000.0) with 3-$\sigma$
statistical uncertainties of $6.\!^{\prime\prime}3$ and $4.\!^{\prime\prime}9$ 
in R.A. and Decl., respectively.  After considering that the systematic 
pointing uncertainty for {\em NuSTAR} is $\sim$$8^{\prime\prime}$ \citep{harrison13}, 
the error region can be approximated with a $10^{\prime\prime}$-radius 
circle.

We searched on-line catalogs (e.g., SIMBAD\footnote{http://simbad.u-strasbg.fr/simbad/})
for X-ray sources consistent with the position of NuSTAR~J163433--4738.7, 
but we did not find any likely candidates.  The catalog from the 2011 
{\em Chandra} survey (Fornasini et al., submitted) does not have any sources 
in the NuSTAR~J163433--4738.7 error region.  Based on this and the analysis
described below (including a re-analysis of the 2011 {\em Chandra} observations), 
we conclude that NuSTAR~J163433--4738.7 is a previously undetected source and 
that it is very likely to be a transient given the sensitivity of the {\em Chandra} 
observations.

\begin{figure}
\includegraphics[scale=0.47]{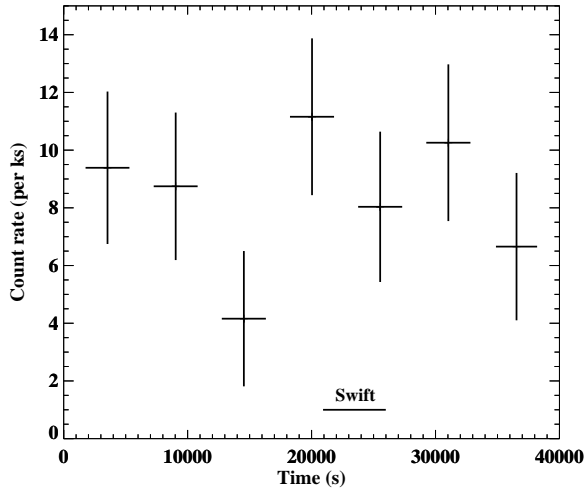}
\caption{\small The 3--10\,keV {\em NuSTAR} light curves (FPMA and FPMB after 
background subtraction) for NuSTAR~J163433--4738.7 during ObsID 40014007001.  
There is one data point per satellite orbit.  The errors on the data points 
are at 1-$\sigma$ confidence.  The time of a {\em Swift} observation is 
indicated.  Zero on the time axis corresponds to MJD~56,346.60000.
\label{fig:lc}}
\end{figure}

For the {\em NuSTAR} and {\em Swift} observations listed in Table~\ref{tab:obs}, 
we estimated count rates or upper limits for NuSTAR~J163433--4738.7 using 
$30^{\prime\prime}$-radius source regions centered at the source position derived 
from ObsID 40014007001.  The XRT 90\% encircled energy radius is approximately 
$20^{\prime\prime}$, but we use $30^{\prime\prime}$ to account for the uncertainty in 
the source position as NuSTAR~J163433--4738.7 is not bright enough in any of 
the {\em Swift} observations to improve the measurement of the source position.  
As described above, we used a larger, circular region that does not contain any 
detected point sources for background.  Selecting the background regions for 
{\em NuSTAR} requires some care because of scattered light from a nearby bright 
source (4U~1630--47).  

The 3--10\,keV count rates or limits obtained for all observations are given 
in Table~\ref{tab:obs}.  For {\em NuSTAR}, the background rates are high enough 
to use Gaussian statistics, and we use the Poisson limits tabulated in \cite{gehrels86} 
for {\em Swift} and {\em Chandra}.  In all cases, we quote the 1-$\sigma$ error 
bars if the minimum of the 1-$\sigma$ error region is positive.  Otherwise, we 
give the 90\% confidence upper limit.  While the {\em NuSTAR} observation taken 
on February 23 provides the only highly significant detection, the {\em Swift} 
observation with the highest count rate is the one that occurred during this 
{\em NuSTAR} observation.  The only other possible evidence for activity from 
NuSTAR~J163433--4738.7 occurred on February 22, when {\em NuSTAR} obtained a 
2.7-$\sigma$ detection in FPMA; however, this is not confirmed by the FPMB data. 

For {\em Chandra} ObsIDs 12529 and 12532 (from 2011) and ObsID 15625 (from 2013),
we analyzed the ACIS data to search for a detection of NuSTAR~J163433--4738.7.
Based on an inspection, no sources are apparent in the 0.3--10\,keV or 3--10\,keV 
images.  The source is $7^{\prime}$ and $5^{\prime}$ from the {\em Chandra} aimpoint 
for the 2011 ObsIDs.  At these off-axis angles, the 90\% encircled energy fraction 
(EEF) radii (for 4.5\,keV photons) are $7.\!^{\prime\prime}4$ and $4.\!^{\prime\prime}5$ 
for ObsIDs 12529 and 12532, respectively.  ObsID 15625 was a dedicated pointing 
with the target on-axis, and the 90\% EEF radius is $2^{\prime\prime}$. For ObsID 12529, 
the largest number of 3--10\,keV counts within any $7.\!^{\prime\prime}4$-radius circle 
inside the {\em NuSTAR} error region is three, and, after accounting for background, 
we calculate a 90\% confidence upper limit of $<$$1.9\times 10^{-4}$\,s$^{-1}$ on 
the count rate (see Table~\ref{tab:obs}).  For ObsID 12532, the largest number of 
3--10\,keV counts within any $4.\!^{\prime\prime}5$-radius circle inside the 
{\em NuSTAR} error region is two, and the count rate limit is 
$<$$2.1\times 10^{-4}$\,s$^{-1}$.  For ObsID 15625, the largest number of 3--10\,keV 
counts within a $2^{\prime\prime}$-radius circle is two, and the count rate limit is
$<$$5.0\times 10^{-4}$\,s$^{-1}$.  This is higher than for the 2011 ObsIDs because
of the lower exposure time.

We made {\em NuSTAR} light curves for ObsID 40014007001 with several different time 
binnings between 0.1\,s and 5500\,s (the approximate satellite orbital period) in 
the 3--10\,keV and 3--79\,keV bandpasses.  The lack of any apparent variability in 
the 0.1--10\,s light curves rules out flares or bursts that might prove the presence 
of a neutron star.  Figure~\ref{fig:lc} shows the 3--10\,keV orbit-by-orbit light 
curve for FPMA and FPMB combined.  At the 5500\,s binning, a $\chi^{2}$ test shows 
that the 3--10\,keV and 3--79\,keV light curves are consistent with the source being 
constant over $\sim$40\,ks.

\begin{figure}
\includegraphics[scale=0.5]{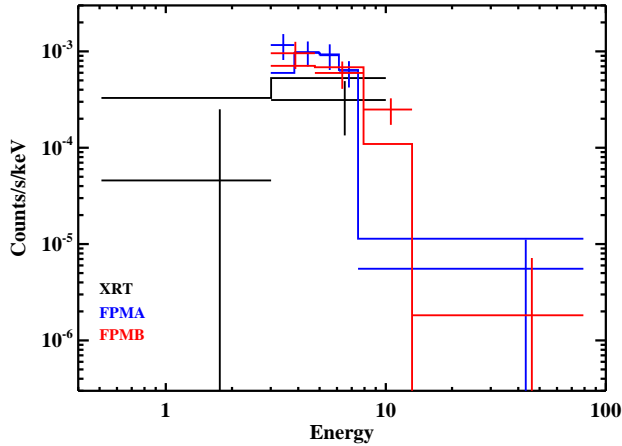}
\caption{\small The {\em NuSTAR} and {\em Swift} energy spectra for 
NuSTAR~J163433--4738.7 on 2013 Feb. 23-24 fitted with an absorbed blackbody 
model.  The black, blue, and red data points and model lines are for XRT, 
FPMA, and FPMB, respectively.  The errors on the data points are 1-$\sigma$ 
confidence.
\label{fig:spectrum}}
\end{figure}

Next, we extracted {\em NuSTAR} FPMA and FPMB spectra for ObsID 40014007001 and an 
XRT spectrum for ObsID 00080508001, which are the two observations with significant
detections of NuSTAR~J163433--4738.7. These were fitted jointly by minimizing the 
Cash (or $C$) statistic \citep{cash79} using the XSPEC software package.  The statistical 
quality of the spectrum is low, and it is well fit by a power-law, a blackbody, or 
a thermal bremsstrahlung model.  In all three cases, we included absorption using 
\cite{wam00} abundances and \cite{vern96} cross-sections.  The parameters are given 
in Table~\ref{tab:spectra}, and the errors quoted are 90\% confidence for one parameter 
of interest, $\Delta$$C$ = 2.7.  Although we used the $C$-statistic to determine the
parameters, we also calculated the reduced-$\chi^{2}$ values, and they appear in 
Table~\ref{tab:spectra}.  This quantity is slightly smaller for the power-law model 
($\chi^{2}_{\nu} = 1.23$ for 8 degrees of freedom) compared to the blackbody model 
($\chi^{2}_{\nu} = 1.44$ for 8 dof) and the bremsstrahlung model 
($\chi^{2}_{\nu} = 1.29$ for 8 dof).  
However, given the small number of dof, the difference in $\chi^{2}_{\nu}$ is not 
significant, and the steeply falling power-law index ($\Gamma = 4.1^{+1.5}_{-1.0}$) 
suggests that the emission probably has a thermal origin.  The spectrum for the 
blackbody model is shown in Figure~\ref{fig:spectrum}.  

\begin{figure}
\includegraphics[scale=0.5]{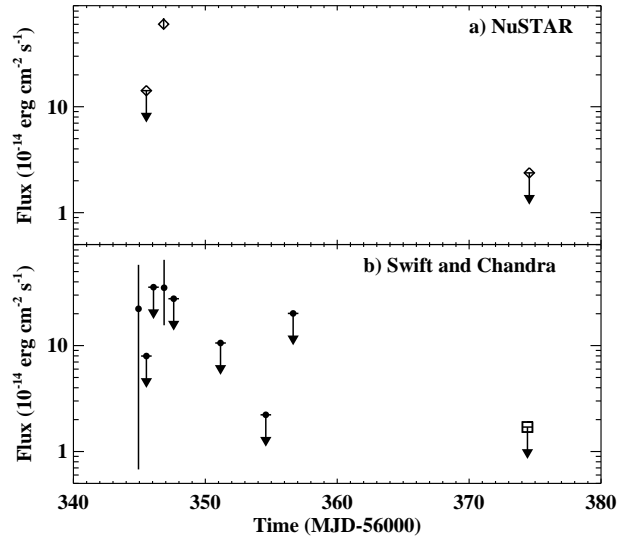}
\caption{\small {\em NuSTAR}, {\em Swift}, and {\em Chandra} (marked with 
diamonds, filled circles, and a square, respectively) absorbed 3--10\,keV
fluxes for NuSTAR~J163433--4738.7 assuming an absorbed blackbody with a 
column density of $N_{\rm H} = 9\times 10^{22}$\,cm$^{-2}$ and a temperature 
of 1.2\,keV.  The errors on the data points are 1-$\sigma$ significance, 
and the upper limits are 90\% confidence.
\label{fig:lc_overview}}
\end{figure}

We used the absorbed blackbody model and the count rates for {\em NuSTAR}, {\em Swift}, 
and {\em Chandra} given in Table~\ref{tab:obs} to calculate measurements or upper 
limits on the absorbed 3--10\,keV flux, and these flux histories are shown in 
Figure~\ref{fig:lc_overview}.  We also calculated the flux upper limits for the 
{\em Chandra} observations from 2011.  The upper limits on the absorbed 3--10\,keV 
fluxes are $<$$1.6\times 10^{-14}$\,erg\,cm$^{-2}$\,s$^{-1}$ and
$<$$2.5\times 10^{-14}$\,erg\,cm$^{-2}$\,s$^{-1}$ for ObsIDs 12529 and 12532, respectively.
Although the count rate limits are considerably lower for these ObsIDs compared
to {\em Chandra} ObsID 15625, the flux limits are similar because of the different
effective areas for the ACIS-I and ACIS-S instruments.  The lowest {\em Chandra}
upper limit indicates that the flux from this source changed by at least a factor
of $39^{+6}_{-4}$, suggesting that it is a transient rather than simply being a
highly variable X-ray source.

The {\em NuSTAR} error circle includes dozens of near-IR candidate 
counterparts brighter than $K_{s}\sim 17$ (Vega magnitude system).  In the 
NEWFIRM images from 2011, the brightest source in the error circle 
(2MASS~J16343288--4738393) has $K_{s} = 12.33\pm 0.05$, $H = 13.07\pm 0.06$, and 
$J = 14.61\pm 0.05$, and the 2MASS magnitudes are consistent, indicating a possible 
lack of variability on long time scales.  The next two brightest sources are 
2MASS~J16343362--4738479 at $K_{s} = 13.31\pm 0.05$ and the {\em Spitzer}/GLIMPSE 
source G336.7870+00.0111 at $K_{s} = 14.70\pm 0.05$.  These may be more likely 
counterparts to NuSTAR~J163433--4738.7 because they are relatively highly reddened 
($H-K_{s} = 2.35\pm 0.08$ and $1.34\pm 0.08$, respectively).  However, we do 
not find any evidence that these sources are variable.  For the two OSIRIS 
$K_{s}$ images taken on April 3 and April 5, we performed aperture photometry 
for all the sources within the {\em NuSTAR} error circle, but did not find any 
variable sources.

The high time-resolution of the {\em NuSTAR} data allows for a search for a 
coherent signal with periods $P \geq 4$\,ms, covering the range expected for 
either an isolated rotation-powered pulsar, a binary, or a magnetar. Given 
the paucity of source counts in the observations listed in Table~1, we 
concentrate our attention on the ObsID 40014007001 data.  Photon arrival times, 
adjusted for the {\em NuSTAR} clock drift, were corrected to the Solar System 
barycenter using JPL DE200 ephemeris and the {\em NuSTAR} derived source 
coordinates. We extracted photons in the 3--10\,keV band from a 
30$^{\prime\prime}$-radius aperture centered on the source to optimize the 
signal-to-noise ratio.  We searched for significant power from a coherent 
signal using an FFT sampled at the Nyquist frequency. The observation span 
was too short to consider the smearing of the pulse profile by spin-down of 
even the most highly energetic pulsar or the typical binary orbit period.
The most significant signal found has a power of $P = 35.18$, corresponding 
to a probability of false detection of $\wp = 0.8$ for $2^{25}$ FFT search 
trials. We conclude that no pulsed X-ray signal is detected in from 
NuSTAR~J163433--4738.7. After taking into account the local background, 
we place an upper limit on the pulse fraction at the 3-$\sigma$ confidence
level of $f_p < 36$\% for a blind search for a sinusoidal signal $P>4$\,ms.

\section{Discussion}

In considering the nature of NuSTAR~J163433--4738.7, it is useful to estimate
the X-ray luminosity that the source reached during its outburst.  Although the 
distance is highly uncertain, the strong absorption either indicates a large 
distance through a region of the Galaxy with heavy extinction or absorption 
local to the source.  The former is a strong possibility since the transient 
is in the direction of a region of the Galactic plane that is crowded with 
H~II/molecular cloud regions.  NuSTAR~J163433--4738.7 is at $l = 336.787^{\circ}, 
b = +0.014^{\circ}$, near a group of H~II regions (336.732+0.072, 336.840+0.047, 
336.9--0.1, 337.147--0.181, and 337.3--0.1) at a distance of $10.9\pm 0.2$\,kpc 
\citep{georgelin96,russeil03}\footnote{This kinematic distance is estimated using
the mean Galactic rotation curve, and the uncertainty does not account for possible
deviations relative to the mean.}.
If the absorption is interstellar, then NuSTAR~J163433--4738.7 is very likely
beyond or in these molecular clouds, indicating a 2--10\,keV unabsorbed 
luminosity limit of $>$$9\times 10^{33}$\,erg~s$^{-1}$ for the blackbody
spectrum and $>$$1.4\times 10^{34}$\,erg~s$^{-1}$ for bremsstrahlung.  In the 
following, we consider both possibilities: a distance of $\sim$11\,kpc and
a peak luminosity of $\sim$$10^{34}$\,erg~s$^{-1}$; and a smaller distance
with absorption local to the source and a lower luminosity.

Considering the first possibility, a luminosity as high as $10^{34}$\,erg~s$^{-1}$ 
would be unprecedented for at least two common types of X-ray sources in the Galaxy.  
Non-magnetic CVs have strong optical outbursts along with persistent and transient 
X-ray emission.  However, extensive studies have shown that their X-ray luminosities 
are in the $10^{29-32}$\,erg~s$^{-1}$ range \citep{baskill05,kuulkers06}.  While 
magnetic CVs can reach higher luminosities \citep{kuulkers06}, they do not typically 
show outbursts.  Secondly, active binaries, including RS~CVn systems and low-mass 
flare stars, produce X-ray flares, some of which can last for about a day.  
Superflares that peak at X-ray luminosities of $10^{32-33}$\,erg~s$^{-1}$ have been 
seen \citep{franciosini01,osten07,osten10,ps12}, and there have been flares that 
have released $\gsim$$10^{37}$\,erg \citep{franciosini01}.  If NuSTAR~J163433--4738.7 
is at $\sim$11\,kpc, then in addition to having a higher peak luminosity, the energy 
released is $\gsim$$4\times 10^{38}$\,erg, which is based on the source being at its
peak luminosity for $\gsim$40\,ks (see Figure~\ref{fig:lc}).  In summary, while we 
do not rule out the active binary possibility (and see below for further discussion 
on this topic), at the 11\,kpc distance, the event seen by {\em NuSTAR} would need 
to be extreme for this explanation to be correct.

A source type that might be a good match to the NuSTAR~J163433--4738.7 properties
is the class of highly magnetic isolated neutron stars: magnetars.  While these 
sources are best known for their very bright and brief ($\sim$0.1\,s) X-ray and 
gamma-ray flares, they also have persistent but variable emission that can easily 
reach $10^{34}$\,erg~s$^{-1}$ or higher \citep{wt06}.  The X-ray spectrum of their 
persistent emission is dominated by a $\sim$0.5--1\,keV blackbody \citep{wt06,mori13}.  
Furthermore, as mentioned in \S\,1, magnetars are associated with regions where 
high-mass star formation is occurring, such as the Norma region, and there is a 
known magnetar, SGR~1627--41, that is $\sim$0.2$^{\circ}$ from NuSTAR~J163433--4738.7.  
However, one possible counter-argument to the magnetar hypothesis is that magnetar 
periods of activity usually last for months.  From Figure~\ref{fig:lc_overview}, 
the peak activity from NuSTAR~J163433--4738.7 could not have lasted for more 
than $\sim$1.5\,days based on the {\em Swift} upper limits, and {\em Chandra} 
and {\em NuSTAR} place a tight upper limit on activity $\sim$3 weeks after 
the {\em NuSTAR} detection.  While we do not detect pulsations or $\sim$0.1\,s
flares, which would prove that the source is a magnetar, the pulsation search 
is limited by the statistical quality of the data, and flaring episodes are 
relatively rare.

A blackbody spectrum could also be produced from an optically thick accretion
disk in a black hole binary.  Most often, such sources show temperatures of
$\sim$1\,keV in the inner parts of their accretion disks when they are at 
high luminosity $\gsim$$10^{36-37}$\,erg~s$^{-1}$, which would require a
very large distance of $\gsim$100--300\,kpc for NuSTAR~J163433--4738.7.  
Two sources (GRS~1758--258 and 1E~1740.7--2942) have shown fainter soft 
states, but their blackbody spectra have been at temperatures of 0.4\,keV 
\citep{smith01} and 0.7\,keV \citep{delsanto05}, which are lower than the 
1.2\,keV we see for NuSTAR~J163433--4738.7.  Also, transient black hole
binaries typically have outbursts that last for several months, which is
very different from NuSTAR~J163433--4738.7.

Considering the second possibility that the source is closer than $\sim$11\,kpc,
then the high column density must be due to material local to the source.  
In fact, {\em INTEGRAL} has found a large number of obscured HMXBs that have
column densities of $10^{23-24}$\,cm$^{-2}$ due to the wind from their supergiant
companions \citep[e.g.][]{walter06}.  While many of the sources in this class
are persistent, the Supergiant Fast X-ray Transients (SFXTs), which have some
members that are also obscured HMXBs, have outbursts that can be as short as 
a few hours, and it would not be unusual for an SFXT to produce X-ray emission 
that lasts for half a day \citep{smith98a,negueruela06,romano11}.  However, a 
possible problem for this interpretation is that the spectrum of 
NuSTAR~J163433--4738.7 is softer than usually seen for active SFXTs.  During 
flares, IGR~J17544--2619 and IGR~J16479--4514 show a power-law spectrum with a 
photon index near $\Gamma = 1.4$ \citep{rampy09,sidoli13}.  Somewhat softer 
spectra with $\Gamma\sim 2.3$--2.4 can be seen during periods of weaker activity, 
but their spectra are only as soft as NuSTAR~J163433--4738.7 when they are 
in quiescence.  

For CVs and active binaries, the challenge is to explain a column density as
high as we measure with absorption local to the systems.  One possibility is
that material expelled from the system (either related to the evolutionary 
state of the stellar component or related to the X-ray flaring event) could
cause increased column density.  However, for one active binary (AX Ari), 
the column density was seen to increase to $N_{\rm H} = 1.1\times 10^{20}$\,cm$^{-2}$
during an event \citep{franciosini01}, which is 2--3 orders of magnitude less 
than is required for NuSTAR~J163433--4738.7.  Variation in optical brightness of 
some stars has also been attributed to ejection of material from the stars.  
An example of this is a change of 0.5 magnitudes in the optical seen for KU Cyg 
\citep{tang11}.  However, this would translate into a column density near 
$10^{21}$\,cm$^{-2}$, which is still far less than is required.

The near-IR information that we have provides only weak constraints on the
source type.  Given that we were not able to identify the counterpart, we can
only say that the source must be fainter (before and after the X-ray flare) 
than the brightest source in the error circle (2MASS~J16343288--4738393).  
This corresponds to $K_{s} > 12.3$, $H > 13.1$, and $J > 14.6$, and these
values are consistent with all the possibilities for the source type discussed
above.  However, in the scenario where NuSTAR~J163433--4738.7 is a nearby
source with local absorption, these limits do provide some constraint.  For
example, a very nearby and luminous obscured HMXB is ruled out.

In summary, NuSTAR~J163433--4738.7 is a fast X-ray transient, which has a thermal
spectrum with relatively high temperature ($kT = 1.2$\,keV for a blackbody or 
3.0\,keV for bremsstrahlung).  If its high column density is due to interstellar
material, the source is probably distant ($\gsim$11\,kpc), making the peak luminosity 
$\gsim$$10^{34}$\,erg\,s$^{-1}$.  We discuss the origin of the flare and suggest that 
the most likely possibilities if the source is distant are an unusually bright flare 
from an active binary or a short outburst from a magnetar.  We also consider
the possibility that the source is closer and that the absorption is local to the
source.  More {\em NuSTAR} observations in the Galactic Plane will determine whether 
such transients are common and, hopefully, shed light on the nature of 
NuSTAR~J163433--4738.7.

\acknowledgments

This work was supported under NASA Contract No. NNG08FD60C, and made use of data 
from the {\it NuSTAR} mission, a project led by  the California Institute of 
Technology, managed by the Jet Propulsion  Laboratory, and funded by the National 
Aeronautics and Space Administration. We thank the {\it NuSTAR} Operations, 
Software and  Calibration teams for support with the execution and analysis of 
these observations.  This research has made use of the {\it NuSTAR}  Data 
Analysis Software (NuSTARDAS) jointly developed by the ASI  Science Data 
Center (ASDC, Italy) and the California Institute of  Technology (USA).
RJA was supported by Gemini-CONICYT grant 32120009.  
FEB was supported by Basal-CATA PFB-06/2007 and CONICYT-Chile (through 
FONDECYT 1101024, Gemini-CONICYT 32120003, and Anillo ACT1101).  
LN wishes to acknowledge the Italian Space Agency (ASI) for financial support by 
ASI/INAF grant I/037/12/0-011/13.  We thank Harvey 
Tananbaum for providing {\em Chandra} Director's Discretionary Time for this project.  
This research has made use of the SIMBAD database, operated at CDS, Strasbourg, France.



\begin{table}
\caption{X-ray Observations and Count Rates for NuSTAR J163433--4738.7\label{tab:obs}}
\begin{minipage}{\linewidth}
\begin{center}
\begin{tabular}{ccccccc} \hline \hline
Mission & Instrument & ObsID & Start Time (UT) & End Time (UT) & Exposure (ks) & Count Rate\footnote{Count rates in the 3--10\,keV band in counts per ks.  The errors given are 1-$\sigma$ and the upper limits are 90\% confidence.}\\ \hline\hline
\multicolumn{7}{c}{Observations in 2011}\\ \hline
{\em Chandra} & ACIS-I & 12529       & Jun 16, 6.97 h   & Jun 16, 12.53 h  & 19.0 & $<$0.19\\
{\em Chandra} & ACIS-I & 12532       & Jun 16, 23.85 h  & Jun 17, 5.55 h   & 19.5 & $<$0.21\\ \hline
\multicolumn{7}{c}{Observations in 2013}\\ \hline
{\em Swift}   & XRT    & 00080509001 & Feb 21, 21.70 h  & Feb 21, 23.13 h  & 1.79 & $1.10^{+1.77}_{-1.06}$\\
{\em NuSTAR}  & FPMA   & 40014004001 & Feb 22, 7.77 h   & Feb 22, 17.52 h  & 19.4 & $1.6\pm 0.6$\\
    ''        & FPMB   &      ''     &       ''         &         ''       &  ''  & $<$1.3\\
{\em Swift}   & XRT    & 00080505001 & Feb 22, 11.88 h  & Feb 22, 13.71 h  & 1.92 & $<$0.54\\
{\em Swift}   & XRT    & 00080506001 & Feb 23, 0.83 h   & Feb 23, 2.62 h   & 1.98 & $<$1.9\\
{\em NuSTAR}  & FPMA   & 40014007001 & Feb 23, 14.52 h  & Feb 24, 1.77 h   & 22.6 & $4.3\pm 0.6$\\
    ''        & FPMB   &      ''     &       ''         &         ''       &  ''  & $4.4\pm 0.9$\\
{\em Swift}   & XRT    & 00080508001 & Feb 23, 20.20 h  & Feb 23, 21.70 h  & 1.97 & $2.2^{+1.8}_{-1.2}$\\
{\em Swift}   & XRT    & 00080511001 & Feb 24, 13.72 h  & Feb 24, 15.18 h  & 1.75 & $<$0.82\\
{\em Swift}   & XRT    & 00032728001 & Feb 28, 1.06 h   & Feb 28, 6.24 h   & 4.95 & $<$0.45\\
{\em Swift}   & XRT    & 00032728002 & Mar 3, 9.00 h    & Mar 3, 19.06 h   & 4.85 & $<$0.12\\
{\em Swift}   & XRT    & 00032728003 & Mar 5, 10.66 h   & Mar 5, 20.75 h   & 4.47 & $<$1.1\\
{\em Chandra} & ACIS-S & 15625       & Mar 23, 8.30 h   & Mar 23, 11.91 h  & 9.84 & $<$0.50\\
{\em NuSTAR}  & FPMA   & 30001012002 & Mar 23, 8.52 h   & Mar 23, 18.35 h  & 16.6 & $<$0.50\\
    ''        & FPMB   &      ''     &      ''          &       ''         &  ''  & $<$2.0\\\hline
\end{tabular}
\end{center}
\end{minipage}
\end{table}

\begin{table}
\caption{Parameters for Fits to {\em NuSTAR} and {\em Swift}/XRT Spectra\label{tab:spectra}}
\begin{minipage}{\linewidth}
\begin{center}
\begin{tabular}{cccccc} \hline \hline
Model & $N_{\rm H}$\footnote{The column density in units of $10^{22}$\,cm$^{-2}$.} & $\Gamma$ or $kT$ & Flux\footnote{The 2--10\,keV unabsorbed flux in units of $10^{-12}$\,erg\,cm$^{-2}$\,s$^{-1}$} & C-statistic\footnote{We fitted the spectra by minimizing the Cash statistic.} & $\chi^{2}_{\nu}$/dof\\ \hline\hline
Power-law & $28^{+23}_{-14}$\footnote{All errors in this table are quoted at the 90\% confidence level.} & $4.1^{+1.5}_{-1.0}$ & $4^{+18}_{-2}$ & 9.5 & 1.23/8\\
Blackbody & $9^{+15}_{-7}$ & $1.2\pm 0.3$\,keV & $1.0^{+0.8}_{-0.3}$ & 11.2 & 1.44/8\\
Bremsstrahlung & $17^{+17}_{-9}$ & $3.0^{+2.1}_{-1.2}$ & $1.6^{+2.0}_{-0.6}$ & 9.8 & 1.29/8\\ \hline
\end{tabular}
\end{center}
\end{minipage}
\end{table}

\end{document}